\documentclass[journal]{IEEEtran}
\usepackage{amsfonts} 
 \usepackage{amssymb}
\usepackage{algorithm}
\usepackage{algpseudocode}

\ifCLASSINFOpdf
   \usepackage[pdftex]{graphicx}
  \else
  \usepackage[dvips]{graphicx}

\fi

\usepackage{amsmath}

\usepackage{array}
\usepackage{subfigure}
\usepackage[2]{editing}

\begin{document}
%
\title{\huge Low-Cost Maximum Entropy Covariance Matrix Reconstruction Algorithm for Robust Adaptive Beamforming}
%
%
%

\author{Saeed~Mohammadzadeh$^{\dag}$, V\'{\i}tor~H.~Nascimento$^{\dag}$ and Rodrigo~C.~de~Lamare$^{\dag\dag}$ 
\thanks{$^\dag$Dept. of Electronic Systems Engineering, Univ. of S\~ao Paulo, Brazil.  $^{\dag\dag}$CETUC, PUC-Rio, Brazil. }}

\maketitle

\begin{abstract}
In this letter, we present a novel low-complexity adaptive beamforming technique using a stochastic gradient algorithm to avoid matrix inversions. The proposed method exploits algorithms based on the maximum entropy power spectrum (MEPS) to estimate the noise-plus-interference covariance matrix (MEPS-NPIC) so that the beamforming weights are updated adaptively, thus greatly reducing the computational complexity. MEPS is further used to reconstruct the desired signal covariance matrix and to improve the estimate of the desired signals's steering vector (SV). Simulations show the superiority of the proposed MEPS-NPIC approach over previously proposed beamformers.
\end{abstract}

%
\IEEEpeerreviewmaketitle

\section{Introduction}
\IEEEPARstart{A}{daptive} beamforming aims to extract the signal from a certain direction while suppressing interference and noise. The technique has been widely used in many fields such as wireless communications, sonar and radar \cite{van2004detection}. However, standard adaptive beamformers are well known to be sensitive to steering vector (SV) mismatches, array imperfections or environmental uncertainties due to non-ideal conditions and many different factors (e.g., wavefront errors, local scattering and finite sample sizes) \cite{li2003robust}. Hence, various adaptive beamformers have been developed in order to mitigate the effects of these problems. Existing methods include diagonal loading \cite{mestre2006finite, kukrer2014generalised,l1stap}, the worst-case optimization in \cite{vorobyov2003robust,wc_ccm} and the projection techniques and subspace-based beamformer investigated in \cite{mohammadzadeh2018adaptive,hassanien2008robust, huang2012modified, shen2015robust,jio,beam_jio,jio_stap,wl_jio,jidf,sjidf,lrcc}. However, these approaches are limited when the SV mismatch is severe. 

Recent works show that the main cause of beamformer performance degradation is the leakage of a component of the signal of interest (SOI) in the sample covariance matrix (SCM) \cite{mallipeddi2011robust}. Recently, a new approach to adaptive beamforming was presented that removes the effect of the SOI component from the covariance matrix by reconstructing the noise-plus-interference covariance (NPIC) matrix. The NPIC matrix in \cite{gu2012robust} is reconstructed by integrating the assumed SV and the corresponding Capon spectrum over a range of angles in which the SOI is absent. The NPIC-based beamformer in \cite{gu2014robust} relies on a sparse reconstruction method. In \cite{ruan2014robust,ruan2016}, computationally efficient algorithms via low complexity reconstruction of the NPIC matrix are presented. In \cite{yuan2017robust} a subspace-based NPIC matrix reconstruction algorithm was proposed. However, it requires that the angular sector of each interference should be known and separated from each other. Later, in \cite{zhang2016interference} an approach is developed using spatial power spectrum sampling (SPSS), but its performance degrades as the number of sensors is decreased. In \cite{chen2018adaptive} the SOI component is eliminated from the DoA of the related bases in order to construct an NPIC matrix directly from the signal-interference subspace.\\
\indent It is worth noting that the use of adaptive antenna arrays and their applications has a trade-off between computational complexity and performance which has a direct relation with the adaptation algorithm.
However, in practice and for large systems, these techniques require the computation of the inverse of the input data SCM (or NPIC matrix), rendering the method very complex.\\
\indent In this letter, we introduce a conjugate gradient (CG) type adaptation algorithm \cite{cg_beam,smcg,l1cg} for the maximum entropy power spectrum noise-plus-interference covariance technique, denoted MEPS-NPIC-CG. The proposed MEPS-NPIC-CG algorithm updates the beamforming weights with a reduced cost as it does not explicitly form the covariance matrices, relying instead on low-cost iterative techniques. The estimated weight vector is obtained from a coarse estimate of the angular sector where the desired signal lies, using CG iterations that avoid the explicit construction of the covariance matrix. We similarly implicitly reconstruct the desired signal covariance matrix and obtain a better desired signal SV estimate using low-cost iterations. An analysis of computational complexity shows that MEPS-NPIC-CG has low-complexity and outperforms other existing techniques.

\section{Problem Background}
Let us assume a linear antenna array with $M$ sensors, spaced by distance $d$, that receive narrowband signals which impinge on the array from several far-field sources. The array observation vector at the $t$-th snapshot can be modeled as
\begin{align}
    \textbf{x}(t)=\textbf{a}(\theta_0) s_0(t)+\sum _{l=1}^L \textbf{a}(\theta_l) s_l(t)+\textbf{n}(t),
\end{align}
where $s_0(t)$ and $s_l(t)$ denote the waveforms of the SOI and $\textit{l}$th interfering signal, respectively. The additive white Gaussian noise vector $\textbf{n}(t)$ is assumed spatially uncorrelated from the desired signal and the interfering signals. The angles $\theta_0$ and $\theta_l$ ($l=1, \cdots, L$) denote the DoAs of the desired signal and interference, respectively. For a sensor array with $\textit{M}$ sensors, $\textbf{a}(\cdot)$ designates the corresponding SV, which has the general form 
    $\textbf{a}(\theta)= \big[1,  e^{-j\pi \bar{d} \sin \theta}, \cdots,  e^{-j\pi(M-1)\bar{d} \sin \theta} \big]^\mathrm{T}$,
where $\bar{d}=2d/\lambda$, $\lambda$ is the wavelength, and $(\cdot)^\mathrm{T}$ denotes the transpose. Assuming that the SV $\textbf{a}_0=\textbf{a}(\theta_0)$ is known, for a given beamformer $\textbf{w}$, the performance is evaluated by the output signal-to-interference-plus-noise ratio (SINR) as 
\begin{align}\label{SINR}
\mathrm{SINR}=\dfrac{\sigma^{2}_0 |\textbf{w}^\mathrm{H} \textbf{a}_0|^2 }{\textbf{w}^\mathrm{H} \textbf{R}_\mathrm{i+n}\textbf{w}},
\end{align}
where $\textbf{R}_\mathrm{i+n}$ is the NPIC matrix, $\sigma^2_0$ is power of the desired signal and $ (\cdot)^\mathrm{H} $ stands for Hermitian transpose. The beamformer that maximizes (\ref{SINR}) is equivalent to finding the solution that maintains a distortionless response toward the SV $\textbf{a}_0$:
\begin{align}\label{MVDR}
\underset{{\textbf{w}}}{\operatorname{min}}\ \textbf{w}^\mathrm{H} \textbf{R}_\mathrm{i+n} \ \textbf{w}\ \hspace{.4cm} \mathrm{s.t.} \hspace{.4cm} \textbf{w}^\mathrm{H} \textbf{a}_0=1.
\end{align}
The solution to (\ref{MVDR}) yields the optimal beamformer given by
\begin{align}\label{optimal wegight vector}
\textbf{w}_{\mathrm{opt}}=\dfrac{\textbf{R}_\mathrm{i+n}^{-1} \textbf{a}_0}{\textbf{a}_0^\mathrm{H} \textbf{R}_\mathrm{i+n}^{-1}\textbf{a}_0},
\end{align}
which is the adaptive weight vector based on the minimum variance distortionless response (MVDR) criterion \cite{van2004detection}. 
Moreover, the array covariance matrix $\textbf{R}=\mathrm{E}\{\textbf{x}(t)\textbf{x}^\mathrm{H}(t)\}$ is 
\begin{equation} \label{Theoretical R}
 \textbf{R}=\textbf{R}_\mathrm{i+n}+\textbf{R}_\mathrm{s}= \int_{\Phi} P(\theta)\textbf{a}(\theta)\textbf{a}^\mathrm{H}(\theta) d\theta,  
\end{equation}
where $P(\theta)$ is the power spectrum of the signals and $\Phi=[\bar{\Theta} \cup \Theta]$ covers the union of the angular sectors of the noise-plus-interference signal, $\bar{\Theta}$, and of the desired signal region, $\Theta$ (obtained through some low-resolution direction finding methods \cite{van2004detection}), while   $\textbf{R}_\mathrm{s}=\sigma_0^2\textbf{a}_0\textbf{a}_0^\mathrm{H}$ is the theoretical desired signal covariance matrix. Since $\textbf{R}_\mathrm{i+n}$ is unknown in practice, it is substituted by the data SCM  \cite{spa} as $\hat{\textbf{R}}=(1/K)\sum_{t=1}^{K} \textbf{x}(t)\textbf{x}^\mathrm{H}(t)$, where $K$ is the number of received snapshot vectors.

\section{Proposed MEPS-NPIC-CG Algorithm}

Our approach is based on MEPS to reconstruct the NPIC matrix and estimate the desired signal SV separately. The NPIC matrix is never explicity computed, however -- instead, we use CG recursions to solve a linear system to obtain the desired weight vector. The desired signal SV is also obtained via a low-complexity approach.
\subsection{Maximum Entropy Power Spectrum }
In the proposed beamforming method, an approach different from prior works is adopted to reconstruct the NPIC and the desired signal covariance matrices. The essence of the idea is based on the use of the spatial spectrum distribution over all possible directions and coarse estimates of the angular regions where the desired signal and the interferers lie. In this work, we exploit maximum entropy power spectrum estimation \cite{lacoss1971data}:
\vspace{-1mm}
\begin{align}\label{Power of MEM}
\hat{{{P}}}_\mathrm{meps}=\dfrac{1}{\epsilon_p  \vert  \textbf{a}^\mathrm{H}(\theta)\hat{\textbf{R}}^{-1} \textbf{u}_1 \vert ^2}
\end{align}
where $\textbf{u}_1=[\begin{smallmatrix}1 & 0 & \cdots & 0\end{smallmatrix}]^\mathrm{T}$,   $\epsilon_p=1/\textbf{u}_1^\mathrm{T}\hat{\textbf{{R}}}^{-1}\textbf{u}_1$.
\subsection{Desired Signal SV Estimation}
In practice, we have inaccurate SV estimates, resulting in performance degradation. Therefore, we utilize the knowledge of the angular sector of the SOI to construct a criterion which can be used to estimate the actual SV. This algorithm is based on the multiplication of an estimate of the desired signal covariance matrix and the nominal SV of the SOI, which results in a vector much closer to the SV of SOI. First, the desired signal covariance matrix can be reconstructed based on MEPS by numerically evaluating \eqref{Theoretical R} over $\Theta$
\begin{align}\label{proposed Rs}
\hat{\textbf{R}}_{\mathrm{s}} \approx \sum_{i=1}^{S} \hat{{{P}}}_\mathrm{meps} \     \textbf{a}(\theta_{s_i})\textbf{a}^\mathrm{H}(\theta_{s_i})\Delta\theta_s,
\end{align}
where $\Theta$ is sampled uniformly with $S$ sampling points spaced by $\Delta\theta_s$, so that $ \{\textbf{a}(\theta_{s_i}) \arrowvert \theta_{s_i} \in \Theta \} $ lies within the range space of $ \hat{\textbf{R}}_{\mathrm{s}} $. Let $\bar{\mathbf{a}}$ be the nominal desired signal SV.  If the set $\Theta$ is such that the noise and interference power are dominated by the signal power in the covariance estimate \eqref{proposed Rs}, then
\begin{align}\label{etimted SV}
\hat{\textbf{a}}_0=\hat{\textbf{R}}_{\mathrm{s}} \bar{\textbf{a}}\simeq (\sigma_\mathrm{0}^2\textbf{a}_0\textbf{a}_0^\mathrm{H})\bar{\textbf{a}}_\simeq \sigma_\mathrm{0}^2(\textbf{a}_0^\mathrm{H}\bar{\textbf{a}})\textbf{a}_0,
\end{align}
 is proportional to the desired signal’s SV (note that the nominal SV is usually a good enough approximation so that  $\mathbf{a}_0^H\bar{\mathbf{a}}$ is far from zero). 
\subsection{NPIC Matrix Reconstruction Using CG}\label{AA}
The classical least mean square (LMS) algorithm in \cite{frost1972algorithm} is based on adjusting the array of sensors in real-time toward a signal coming from the desired direction while the interferences are attenuated. The LMS algorithm is an CG algorithm which searches for the minimum of a quadratic cost function. We apply LMS to solve the MVDR optimization problem in (\ref{MVDR}) by using the Lagrange multiplier $\alpha$ to include the constraint into the objective function as 
\begin{align} \label{cost function}
    J(\textbf{w})=\textbf{w}^\mathrm{H} \hat{\textbf{R}}_\mathrm{i+n} \ \textbf{w}+\alpha(\textbf{w}^\mathrm{H} \hat{\textbf{a}}_0-1).
\end{align}
The cost function $J(\textbf{w})$  can be minimized by applying the steepest descent algorithm  as follows 
\begin{align} \label{Iteration}
  \textbf{w}(t+1)=\textbf{w}(t)-\dfrac{1}{2} \mu \nabla  J(\textbf{w}),
\end{align}
where $\nabla  J(\textbf{w})$ is the gradient of the cost function with respect to $\textbf{w}(t)$. The gradient vector can be obtained from (\ref{cost function}) as
\begin{align} \label{Gradient}
  \nabla  J(\textbf{w})=2 \hat{\textbf{R}}_\mathrm{i+n} \textbf{w}(t)+\alpha \hat{\textbf{a}}_0. 
\end{align}
Exploiting \eqref{Theoretical R} over angular $\bar{\Theta}$ and the MEPS estimate (\ref{Power of MEM}), the NPIC matrix can be reconstructed by numerically evaluating
\begin{align}\label{proposed Ri+n}
\hat{\textbf{R}}_{\mathrm{i+n}}=\int_{\bar{\Theta}}\hat{{{P}}}_\mathrm{meps}\textbf{a}(\theta)\textbf{a}^\mathrm{H}(\theta)d\theta,
\end{align}
Sampling $\bar{\Theta}$ uniformly with $Q$ sampling points 
spaced by $\Delta\theta$, (\ref{proposed Ri+n}) can be approximated by
\begin{align} \label{Summation}
\hat{\textbf{R}}_{\mathrm{i+n}} \approx \sum_{i=1}^{Q} \dfrac{\textbf{a}(\theta_i)\textbf{a}^\mathrm{H}(\theta_i)}
{\epsilon_p  \vert  \textbf{a}^\mathrm{H}(\theta)\hat{\textbf{R}}^{-1} \textbf{u}_1 \vert ^2} \Delta\theta.
\end{align}
Here and in the next section we show how to apply the update \eqref{Iteration} while avoiding to compute \eqref{Summation} explicitly.
Rewriting (\ref{Gradient}) by substituting the expression for $\hat{\textbf{R}}_{\mathrm{i+n}}$, we get
\begin{align} \label{Grad with w}
 \nabla  J(\textbf{w}) = 
  \sum_{i=1}^{Q} 2\hat{{{P}}}_\mathrm{meps} \Big(\textbf{a}^\mathrm{H}(\theta_i)   \textbf{w}(t)\Big)\textbf{a}(\theta_i)\Delta\theta +\alpha \hat{\textbf{a}}_0, 
\end{align}
By substituting (\ref{Grad with w}) into (\ref{Iteration}) and rearranging, we obtain a recursion for the beamformer  given by
\begin{align} \label{final W}
  \textbf{w}(t+1)=\textbf{w}(t)-\mu \big[ \hat{\textbf{a}}_0- \textbf{r}(t) \big],
\end{align}
where $\textbf{r}(t)=\sum_{i=1}^{Q}\hat{{{P}}}_\mathrm{meps} \Big(\textbf{a}^\mathrm{H}(\theta_i)   \textbf{w}(t)\Big)\textbf{a}(\theta_i)\Delta\theta$ and $\mu$ is the steepest descent step size. In order to find the beamformer, the conjugate gradient algorithm is used to solve the unconstrained
quadratic programming problem in \eqref{cost function} as algorithm 1.
\begin{algorithm}
\caption{Conjugate Gradient \cite{luenberger1984linear}}
\begin{algorithmic}[1]
\State Choose an initial iterate $\textbf{w}_0$;
\State Set $\textbf{g}_0= \nabla  J(\textbf{w}_0)$ and 
$\textbf{e}_0=-\textbf{g}_0$;
\State Set $t \gets 0$;
\While{$\Vert \nabla  J(\textbf{w}_t) \Vert > \mathrm{tol.}$}
\State Define $\textbf{e}_t=\hat{\textbf{a}}_0- \textbf{r}_t$
\State Determine the step-size $\mu_t=-\dfrac{\textbf{e}_t^\mathrm{T} \textbf{g}_t }{\textbf{e}_t^\mathrm{T}\hat{\textbf{R}}_{\mathrm{i+n}} \textbf{e}_t }$ 
\State $\textbf{w}_{t+1}=\textbf{w}_t+\mu_t \textbf{e}_t$;
\State $\textbf{g}_{t+1}=\nabla  J(\textbf{w}_{t+1})$
\State Determine $\beta_t$  as $\beta_t=\dfrac{\textbf{g}_{t+1}^\mathrm{T} (\textbf{g}_{t+1}-\textbf{g}_{t})}{\textbf{g}_{t}^\mathrm{T} \textbf{g}_{t}}$;
\State Set $\textbf{e}_{t+1}=-\textbf{g}_{t+1}+\beta_t\textbf{e}_t$;
\State Set $t \gets t+1$;
\EndWhile
\end{algorithmic}
\end{algorithm}
%
Hence, the weight vector is updated at each iteration by the recursion in (\ref{final W}) for reducing the complexity.\\
Up to now, the main difference here from prior works lies in the fact that the integral (\ref{proposed Ri+n}) is approximated by a summation (\ref{Summation}), which would require a complexity of $ \mathcal{O}(M^2Q)$ to be able to synthesize narrowband signal's power accurately. However, in the computation of \eqref{Grad with w} and in the final proposed algorithm in (\ref{final W}),  \eqref{V recursion} and \eqref{xi.approx} (see below), we avoid actually computing expensive $O(M^2)$ outer products, so our algorithm  requires $\mathcal{O}(MQ)$ for steps (2,4,5,7,8 and 10) while steps in (6 and 9) needs $\mathcal{O}(M)$ complexity.  Since this algorithm iterates t times to finding the best step-size, $\mu_t$. Hence, the final computational complexity of the proposed method is only  $\mathcal{O}(tMQ)$  while  computing the beamformer without need for the inverse of the NPIC matrix.
\subsection{MEPS Estimation Using CG}\label{sec:MEPS.CG}
In order to compute \eqref{Summation} efficiently, we can use an iterative solution to the linear system, and take advantage of the structure of the SCM, $\hat{\textbf{R}}$. We write the term $\textbf{v}=\hat{\textbf{R}}^{-1} \textbf{u}_1$ in \eqref{Summation} and consider the optimization problem 
\begin{align}\label{V}
\underset{{\textbf{v}}}{\operatorname{min}}\ \textbf{v}^\mathrm{H} \hat{\textbf{R}} \ \textbf{v}\ \hspace{.4cm} \mathrm{s.t.} \hspace{.4cm} \textbf{v}^\mathrm{H} \textbf{u}_1=1,
\end{align}
The corresponding CG algorithm is described by
\begin{align} \label{V rec}
    \textbf{v}(t+1)=\textbf{v}(t)+ \xi (\textbf{u}_1-\hat{\textbf{R}}\textbf{v}(t)),
\end{align}
where $\xi$ is a step size. 
Now, substituting the expression for $\hat{\textbf{R}}$ and multiplying by $\textbf{v}(t)$ yields
\begin{align} \label{R hat with v}
    \hat{\textbf{R}}\textbf{v}(t)=\dfrac{1}{K}\sum_{t=1}^K \textbf{x}(t)\Big(\textbf{x}^\mathrm{H}(t)\textbf{v}(t) \Big),
\end{align}
By substituting (\ref{R hat with v}) into (\ref{V rec}), we obtain
\begin{align} \label{V recursion}
    \textbf{v}(t+1)=\textbf{v}(t)+ \xi \Big(\textbf{u}_1- \sum_{t=1}^K \textbf{x}(t) \big( \dfrac{\textbf{x}^\mathrm{H}(t)\textbf{v}(t)}{K}\big) \Big).
\end{align}
In (\ref{V rec}), the step size, $\xi$, should satisfy $0 < \xi < 2/\lambda_\mathrm{max}$ ($\lambda_{\mathrm{max}}$ is the largest eigenvalue of $\hat{\mathbf{R}}$), with fastest convergence occurring for $\xi \approx 1/\lambda_\mathrm{max}$. Since computing $\lambda_\mathrm{max}$  requires $ \mathcal{O}(M^3)$ operations, it is more efficient to use an approximation.
Assume that $\lambda$ is an eigenvalue of $\hat{\textbf{R}}$ with respect to the eigenvalue $\textbf{z}$, so we can write
\begin{align}
    \lambda \textbf{z}=\hat{\textbf{R}}\textbf{z}=\dfrac{1}{K}\sum_{t=1}^K \textbf{x}(t)\Big(\textbf{x}^\mathrm{H}(t)\textbf{z} \Big).
\end{align}
Taking norm in both sides
\begin{multline}
  \mid \lambda \mid  \parallel \textbf{z} \parallel  = \dfrac{1}{K} \parallel \sum_{t=1}^K \textbf{x}(t)\textbf{x}^\mathrm{H}(t)\textbf{z}  \parallel \\ \leqslant \dfrac{1}{K} \sum_{t=1}^K \parallel \textbf{x}(t) \parallel \mid \textbf{x}^\mathrm{H}(t)\textbf{z}\mid 
    \leqslant \dfrac{1}{K} \sum_{t=1}^K \parallel \textbf{x}(t) \parallel^2 \parallel \textbf{z} \parallel, 
  \end{multline}
Hence  $\mid \lambda \mid \leqslant (1/K) \sum _{t=1}^K \parallel \textbf{x}(t) \parallel ^2 $.
An approximation to the step size, $\xi$, is given by
\begin{align}\label{xi.approx}
    \xi \approx \dfrac{K}{\sum _{t=1}^K \parallel \textbf{x}(t) \parallel ^2}.
\end{align}
The computational complexity of MEPS-NPIC-CG is $ \mathcal{O}(tMQ)$. The solution of the quadratically constrained quadratic programming (QCQP) problem in \cite{gu2012robust} has complexity of at least $ \mathcal{O}(M^{3.5}+M^2Q) $,  while the beamformer in \cite{huang2012modified} has a complexity of $ \mathcal{O}(KM)+\mathcal{O}(M^{3}) $ and the reconstructed NPIC matrices in \cite{ruan2014robust} and \cite{zhang2016interference} have a complexity of $ \mathcal{O}(M^3) $. 
Also, the cost of the beamformer in \cite{zheng2018covariance} is $ \mathcal{O}(\mathrm{max}(M^2Q,M^{3.5})) $.  
\section{Simulations}
In this section, a uniform linear array of $M=10$ omnidirectional sensors and half-wavelength interelement spacing is considered. Two interferers and a desired signal impinge on the sensor array with incident angles $50^o$, $20^o$ and $5^o$, respectively. The interference-plus-noise ratio (INR) for each interferer is assumed 30 dB in each sensor. The additive noise is modeled as spatially white Gaussian with zero mean and unit variance where 100 Monte Carlo runs are performed for each simulation. When we examine the performance of the output SINR versus input SNR, the number of snapshots is set to $K = 30$ whereas for the performance comparison of the adaptive beamformers versus the number of snapshots the SNR is set to 20 dB.
The proposed MEPS-NPIC-CG method is compared with  LOCSME \cite{ruan2014robust}, the modified projection beamformer (Shrinkage) \cite{huang2012modified}, the reconstruction-estimation based beamformer (Rec-Est) \cite{gu2012robust}, the SPSS beamformer in \cite{zhang2016interference}, 
the algorithm based on noise-plus-interference covariance matrix reconstruction and SV estimation (INC-SV), \cite{zheng2018covariance} and the beamformer (SV-Est) in \cite{khabbazibasmenj2012robust}. The angular sector of the desired signal is set to $ \Theta=[-1^\circ,11^\circ] $ while the interference angular sector is $ \bar{\Theta}=[-90^\circ,-1^\circ)\cup(11^\circ,90^\circ] $. For the proposed MEPS-NPIC-CG beamformer, $\mathrm{tol}=0.001$, $t=7$, $S=10$ and $Q=90$ are used and the bound for the beamformer in \cite{zheng2018covariance} is set as $ \epsilon=\sqrt{0.1} $. To solve the optimization problems to compute the optimum solutions used for comparison, the Matlab CVX toolbox \cite{grant2017cvx} is used.

\begin{figure}[!] \label{SNR}
  \subfigure{\includegraphics[width=0.22\textwidth]{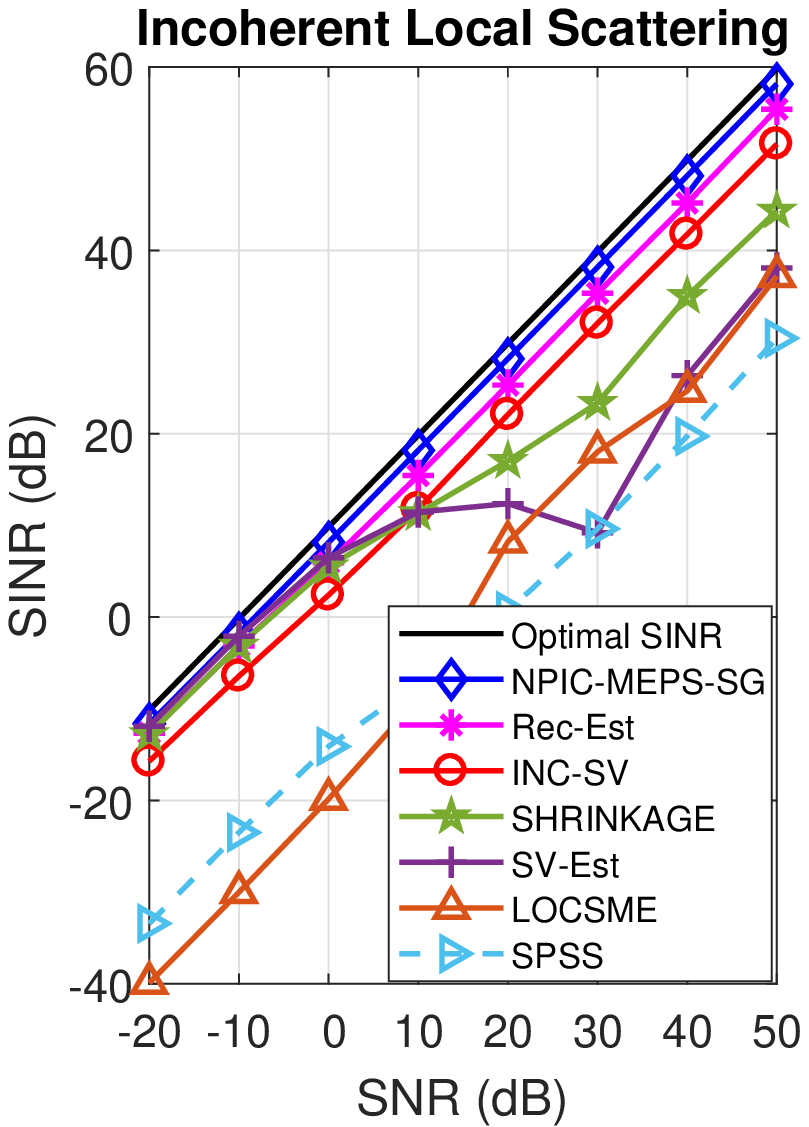}\label{SINR_SNRInco}}\hfill
  \subfigure{\includegraphics[width=0.22\textwidth]{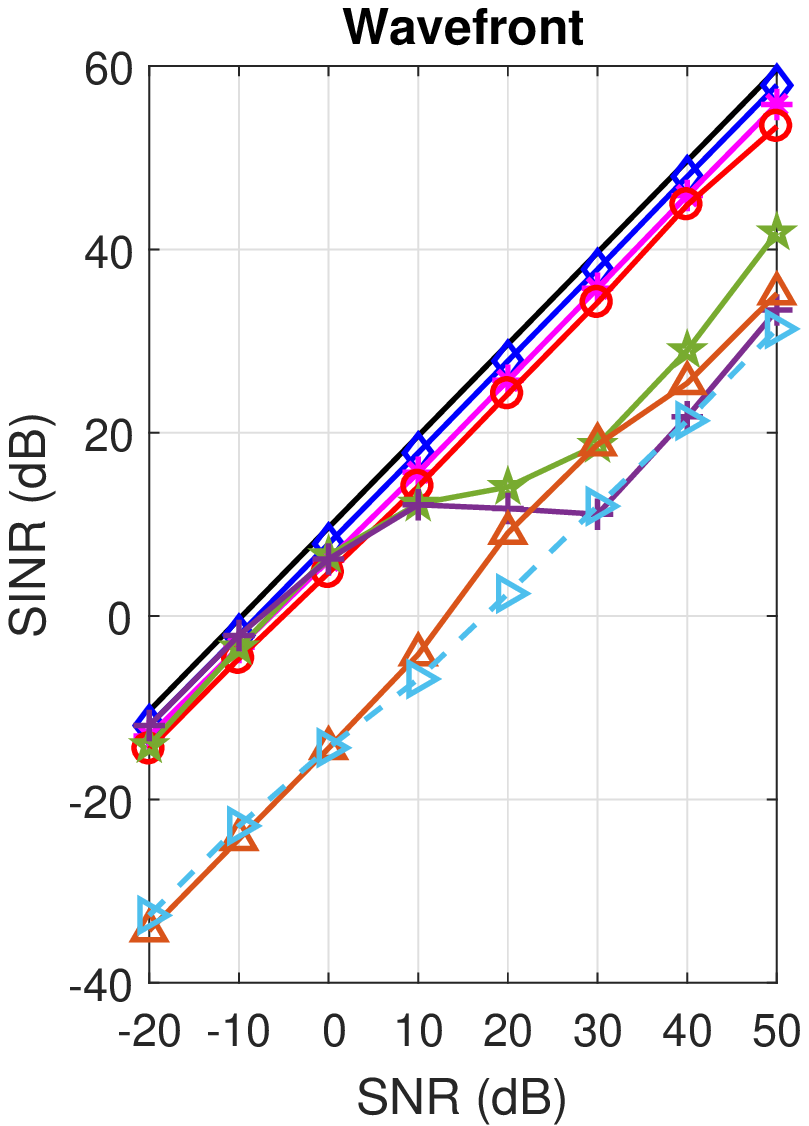}\label{SINR_SNRWave}}
  \vspace{-0.85em}
  \caption{SINR vs SNR \ a) Incoherent Local Scattering \ b) Wavefront }
\end{figure}

\begin{figure}[!] \label{Snapshots}
  \subfigure{\includegraphics[width=0.22\textwidth]{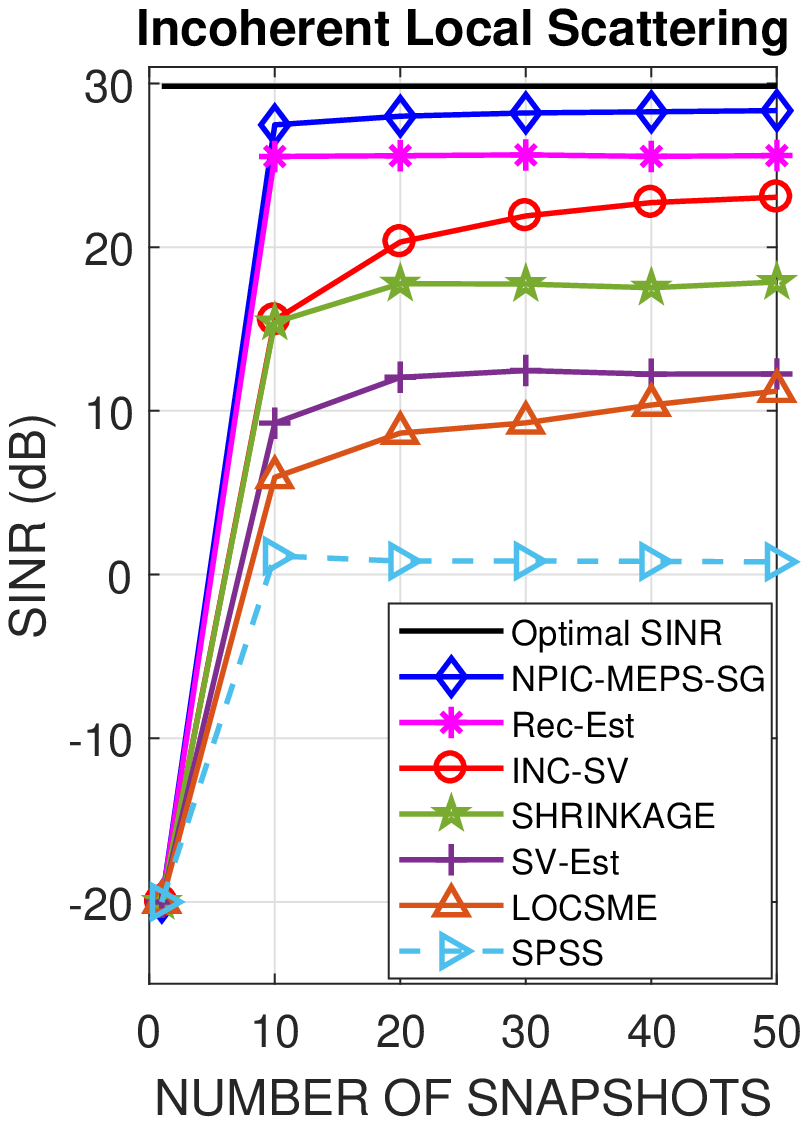}\label{SINR_SNInco}}\hfill
  \subfigure{\includegraphics[width=0.22\textwidth]{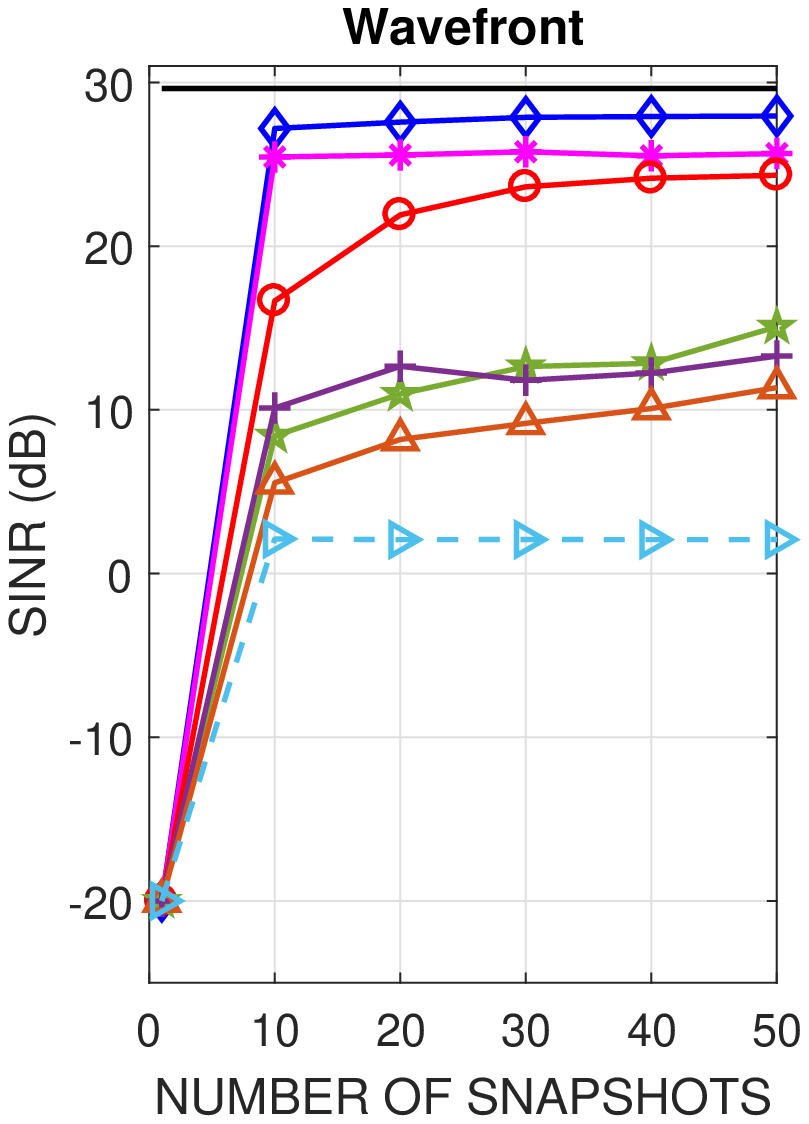}\label{SINR_SNWave}}
  \vspace{-0.85em}
  \caption{SINR vs Snapshots \ a) Incoherent Local Scattering \ b)  Wavefront}\label{fig:snapshots}
  \end{figure}

In the first scenario, we investigate the impact of the array arbitrary imperfection mismatch in an inhomogeneous medium when the wave propagation effects distort the signal spatial signature. Specifically, when the nominal SV components accumulate the phase increment, it is assumed that the phase increments are chosen independently from the random error vector of the Gaussian generator with zero mean and standard deviation 0.07, which in each simulation run remain fixed. 
Figs.~\ref{SINR_SNRWave} and Figs.~\ref{SINR_SNWave} show the output SINR of the beamformers versus the input SNR and versus the snapshots. Similar to the previous scenario, the proposed beamformer significantly outperforms other beamformers due to its ability to reconstruct the NPIC matrix and estimate the desired signal SV with higher accuracy than other methods.


In the second scenario, we analyze the effect of the SV mismatch by the incoherent local scattering of the desired signal. We assume that the actual SV has a time-varying signature and the SV is expressed as $ \textbf{a}(t)=s_0(t)\textbf{a}(\theta_0)+\sum_{p=1}^4 s_p(t)\textbf{a}(\theta_p)$
where $s_0(t)$ and $s_p(t)$ $(p=1,2,3,4)$ are independently and identically distributed complex Gaussian random variables independently drawn from a random generator. In each simulation run, the DoAs $\theta_p$ $(p=0,1,2,3,4)$ are independently distributed in a uniform generator with mean $5^\circ$ and standard deviation $2^\circ$. Note that $\theta_p$ changes from run to run while remains fixed from snapshot to snapshot. Meanwhile, the random variables $s_0(t)$ and $s_p(t)$ change from both run to run and snapshot to snapshot. Fig.~\ref{SINR_SNRInco} and Fig.~\ref{SINR_SNInco} depict the output SINR of the tested beamformers versus the SNR and versus the snapshots under the incoherent local scattering case. It is demonstrated that the MEPS-NPIC-CG has high accuracy SINR for all snapshots. Also, it is seen that the result of the proposed (MEPS-NPIC-CG) beamformer outperforms the other beamformers.
\section{Conclusion}
A low-complexity approach to robust adaptive beamforming based on estimated weight vector through CG recursions, named MEPS-NPIC-CG, has been proposed. The computed weight vector is exploited  to reconstruct accurate NPIC matrix without requiring matrix inversions. Simulations demonstrate that MEPS-NPIC-CG can offer a superior performance to recently reported robust adaptive beamforming methods.

\ifCLASSOPTIONcaptionsoff
\fi
\bibliographystyle{IEEEtran}
\bibliography{document}

\begin{thebibliography}{10}
\providecommand{\url}[1]{#1}
\csname url@samestyle\endcsname
\providecommand{\newblock}{\relax}
\providecommand{\bibinfo}[2]{#2}
\providecommand{\BIBentrySTDinterwordspacing}{\spaceskip=0pt\relax}
\providecommand{\BIBentryALTinterwordstretchfactor}{4}
\providecommand{\BIBentryALTinterwordspacing}{\spaceskip=\fontdimen2\font plus
\BIBentryALTinterwordstretchfactor\fontdimen3\font minus
  \fontdimen4\font\relax}
\providecommand{\BIBforeignlanguage}[2]{{%
\expandafter\ifx\csname l@#1\endcsname\relax
\typeout{** WARNING: IEEEtran.bst: No hyphenation pattern has been}%
\typeout{** loaded for the language `#1'. Using the pattern for}%
\typeout{** the default language instead.}%
\else
\language=\csname l@#1\endcsname
\fi
#2}}
\providecommand{\BIBdecl}{\relax}
\BIBdecl

\bibitem{van2004detection}
H.~L. Van~Trees, \emph{Detection, Estimation, and Modulation Theory}.\hskip 1em
  plus 0.5em minus 0.4em\relax John Wiley \& Sons, New York, 2004.

\bibitem{li2003robust}
J.~Li, P.~Stoica, and Z.~Wang, ``On robust {Capon} beamforming and diagonal
  loading,'' \emph{IEEE Trans. on Signal Process.}, vol.~51, no.~7, pp.
  1702--1715, 2003.

\bibitem{mestre2006finite}
X.~Mestre and M.~A. Lagunas, ``Finite sample size effect on minimum variance
  beamformers: Optimum diagonal loading factor for large arrays,'' \emph{IEEE
  Trans. on Signal Process.}, vol.~54, no.~1, pp. 69--82, 2006.

\bibitem{kukrer2014generalised}
O.~Kukrer and S.~Mohammadzadeh, ``Generalised loading algorithm for adaptive
  beamforming in ulas,'' \emph{Electronics Letters}, vol.~50, no.~13, pp.
  910--912, 2014.

\bibitem{l1stap}
Z.~{Yang}, R.~C. {de Lamare}, and X.~{Li}, ``$l_1$ -regularized stap algorithms
  with a generalized sidelobe canceler architecture for airborne radar,''
  \emph{IEEE Transactions on Signal Processing}, vol.~60, no.~2, pp. 674--686,
  2012.

\bibitem{vorobyov2003robust}
S.~A. Vorobyov, A.~B. Gershman, and Z.-Q. Luo, ``Robust adaptive beamforming
  using worst-case performance optimization: A solution to the signal mismatch
  problem,'' \emph{IEEE Trans. on Signal Process.}, vol.~51, no.~2, pp.
  313--324, 2003.

\bibitem{wc_ccm}
L.~{Landau}, R.~C. {de Lamare}, and M.~{Haardt}, ``Robust adaptive beamforming
  algorithms using the constrained constant modulus criterion,'' \emph{IET
  Signal Processing}, vol.~8, no.~5, pp. 447--457, 2014.

\bibitem{mohammadzadeh2018adaptive}
S.~Mohammadzadeh and O.~Kukrer, ``Adaptive beamforming based on theoretical
  interference-plus-noise covariance and direction-of-arrival estimation,''
  \emph{IET Signal Process.}, vol.~12, no.~7, pp. 819--825, 2018.

\bibitem{hassanien2008robust}
A.~Hassanien, S.~A. Vorobyov, and K.~M. Wong, ``Robust adaptive beamforming
  using sequential quadratic programming: An iterative solution to the mismatch
  problem,'' \emph{IEEE Signal Process.Lett.}, vol.~15, pp. 733--736, 2008.

\bibitem{huang2012modified}
F.~Huang, W.~Sheng, and X.~Ma, ``Modified projection approach for robust
  adaptive array beamforming,'' \emph{Signal Process.}, vol.~92, no.~7, pp.
  1758--1763, 2012.

\bibitem{shen2015robust}
F.~Shen, F.~Chen, and J.~Song, ``Robust adaptive beamforming based on steering
  vector estimation and covariance matrix reconstruction,'' \emph{IEEE
  Communications Letters}, vol.~19, no.~9, pp. 1636--1639, 2015.

\bibitem{jio}
R.~C. {de Lamare} and R.~{Sampaio-Neto}, ``Reduced-rank adaptive filtering
  based on joint iterative optimization of adaptive filters,'' \emph{IEEE
  Signal Processing Letters}, vol.~14, no.~12, pp. 980--983, 2007.

\bibitem{beam_jio}
``Adaptive reduced-rank lcmv beamforming algorithms based on joint iterative
  optimization of filters: Design and analysis,'' \emph{Signal Processing},
  vol.~90, no.~2, pp. 640 -- 652, 2010.

\bibitem{jio_stap}
R.~{Fa} and R.~C. {De Lamare}, ``Reduced-rank stap algorithms using joint
  iterative optimization of filters,'' \emph{IEEE Transactions on Aerospace and
  Electronic Systems}, vol.~47, no.~3, pp. 1668--1684, 2011.

\bibitem{wl_jio}
N.~{Song}, W.~U. {Alokozai}, R.~C. {de Lamare}, and M.~{Haardt}, ``Adaptive
  widely linear reduced-rank beamforming based on joint iterative
  optimization,'' \emph{IEEE Signal Processing Letters}, vol.~21, no.~3, pp.
  265--269, 2014.

\bibitem{jidf}
R.~C. {de Lamare} and R.~{Sampaio-Neto}, ``Adaptive reduced-rank processing
  based on joint and iterative interpolation, decimation, and filtering,''
  \emph{IEEE Transactions on Signal Processing}, vol.~57, no.~7, pp.
  2503--2514, 2009.

\bibitem{sjidf}
R.~{Fa}, R.~C. {de Lamare}, and L.~{Wang}, ``Reduced-rank stap schemes for
  airborne radar based on switched joint interpolation, decimation and
  filtering algorithm,'' \emph{IEEE Transactions on Signal Processing},
  vol.~58, no.~8, pp. 4182--4194, 2010.

\bibitem{lrcc}
H.~{Ruan} and R.~C. {de Lamare}, ``Distributed robust beamforming based on
  low-rank and cross-correlation techniques: Design and analysis,'' \emph{IEEE
  Transactions on Signal Processing}, vol.~67, no.~24, pp. 6411--6423, 2019.

\bibitem{mallipeddi2011robust}
R.~Mallipeddi, J.~P. Lie, S.~G. Razul, P.~Suganthan, and C.~M.~S. See, ``Robust
  adaptive beamforming based on covariance matrix reconstruction for look
  direction mismatch,'' \emph{Progress In Electromagnetics Research}, vol.~25,
  pp. 37--46, 2011.

\bibitem{gu2012robust}
Y.~Gu and A.~Leshem, ``Robust adaptive beamforming based on interference
  covariance matrix reconstruction and steering vector estimation,'' \emph{IEEE
  Trans. on Signal Process.}, vol.~60, no.~7, pp. 3881--3885, 2012.

\bibitem{gu2014robust}
Y.~Gu, N.~A. Goodman, S.~Hong, and Y.~Li, ``Robust adaptive beamforming based
  on interference covariance matrix sparse reconstruction,'' \emph{Signal
  Processing}, vol.~96, pp. 375--381, 2014.

\bibitem{ruan2014robust}
H.~Ruan and R.~C. de~Lamare, ``Robust adaptive beamforming using a
  low-complexity shrinkage-based mismatch estimation algorithm.'' \emph{IEEE
  Signal Process. Lett.}, vol.~21, no.~1, pp. 60--64, 2014.

\bibitem{ruan2016}
H.~{Ruan} and R.~C. {de Lamare}, ``Robust adaptive beamforming based on
  low-rank and cross-correlation techniques,'' \emph{IEEE Trans. on Signal
  Process.}, vol.~64, no.~15, pp. 3919--3932, Aug 2016.

\bibitem{yuan2017robust}
X.~Yuan and L.~Gan, ``Robust adaptive beamforming via a novel subspace method
  for interference covariance matrix reconstruction,'' \emph{Signal
  Processing}, vol. 130, pp. 233--242, 2017.

\bibitem{zhang2016interference}
Z.~Zhang, W.~Liu, W.~Leng, A.~Wang, and H.~Shi, ``Interference-plus-noise
  covariance matrix reconstruction via spatial power spectrum sampling for
  robust adaptive beamforming,'' \emph{IEEE Signal Process. Lett.}, vol.~23,
  no.~1, pp. 121--125, 2016.

\bibitem{chen2018adaptive}
P.~Chen, Y.~Yang, Y.~Wang, and Y.~Ma, ``Adaptive beamforming with sensor
  position errors using covariance matrix construction based on subspace bases
  transition,'' \emph{IEEE Signal Process. Lett.}, vol.~26, no.~1, pp. 19--23,
  2018.

\bibitem{cg_beam}
L.~{Wang} and R.~C.~D. {Lamare}, ``Constrained adaptive filtering algorithms
  based on conjugate gradient techniques for beamforming,'' \emph{IET Signal
  Processing}, vol.~4, no.~6, pp. 686--697, 2010.

\bibitem{smcg}
L.~{Wang} and R.~C. {de Lamare}, ``Set-membership constrained conjugate
  gradient adaptive algorithm for beamforming,'' \emph{IET Signal Processing},
  vol.~6, no.~8, pp. 789--797, 2012.

\bibitem{l1cg}
Z.~{Yang}, R.~C.~D. {Lamare}, and X.~{Li}, ``Sparsity-aware space-time adaptive
  processing algorithms with l1-norm regularisation for airborne radar,''
  \emph{IET Signal Processing}, vol.~6, no.~5, pp. 413--423, 2012.

\bibitem{spa}
R.~C. {De Lamare} and R.~{Sampaio-Neto}, ``Minimum mean-squared error iterative
  successive parallel arbitrated decision feedback detectors for ds-cdma
  systems,'' \emph{IEEE Transactions on Communications}, vol.~56, no.~5, pp.
  778--789, 2008.

\bibitem{lacoss1971data}
R.~T. Lacoss, ``Data adaptive spectral analysis methods,'' \emph{Geophysics},
  vol.~36, no.~4, pp. 661--675, 1971.

\bibitem{frost1972algorithm}
O.~L. Frost, ``An algorithm for linearly constrained adaptive array
  processing,'' \emph{Proceedings of the IEEE}, vol.~60, no.~8, pp. 926--935,
  1972.

\bibitem{luenberger1984linear}
D.~G. Luenberger, Y.~Ye \emph{et~al.}, \emph{Linear and nonlinear
  programming}.\hskip 1em plus 0.5em minus 0.4em\relax Springer, 1984, vol.~2.

\bibitem{zheng2018covariance}
Z.~Zheng, Y.~Zheng, W.-Q. Wang, and H.~Zhang, ``Covariance matrix
  reconstruction with interference steering vector and power estimation for
  robust adaptive beamforming,'' \emph{IEEE Trans. on Vehicular Tech.},
  vol.~67, no.~9, pp. 8495--8503, 2018.

\bibitem{khabbazibasmenj2012robust}
A.~Khabbazibasmenj, S.~A. Vorobyov, and A.~Hassanien, ``Robust adaptive
  beamforming based on steering vector estimation with as little as possible
  prior information,'' \emph{IEEE Trans. on Signal Process.}, vol.~60, no.~6,
  pp. 2974--2987, 2012.

\bibitem{grant2017cvx}
M.~Grant and S.~Boyd, ``Cvx: Matlab software for disciplined convex
  programming, version 2.1,'' 2017.

\end{thebibliography}
\end{document}